\documentclass{PoS}

\usepackage{psfrag}
\usepackage{color}
\usepackage{amsmath, amsthm, amssymb, bm}

\newcommand{\mygraph}[1]{\includegraphics[width=#1,keepaspectratio]}
\newcommand{\Red}{\textcolor{red}}
\newcommand{\Blue}{\textcolor{blue}}
\newcommand{\msub}[1]{\ensuremath _{\mbox{\tiny #1}}}
\newcommand{\be}{\begin{equation}}
\newcommand{\ee}{\end{equation}}

\title{Dirac eigenmodes at the QCD Anderson transition}

\ShortTitle{Dirac eigenmodes at the QCD Anderson transition}

\author{Matteo Giordano\thanks{Supported by the Hungarian Academy of Sciences
    under ``Lend\"ulet'' grant No.\ LP2011-011.} \\ Institute for Nuclear
  Research of the Hungarian Academy of Sciences \\ Bem t\'er 18/c H-4026
  Debrecen, Hungary \\ E-mail: \email{kgt@atomki.mta.hu}}

\author{\speaker{Tamas G. Kovacs}\footnotemark[1] \\
   Institute for Nuclear Research of the Hungarian Academy of Sciences \\ 
   Bem t\'er 18/c H-4026 Debrecen,  Hungary \\ 
   E-mail: \email{kgt@atomki.mta.hu}}

\author{Ferenc Pittler \\
        MTA-ELTE Lattice Gauge Theory Research Group \\
        P\'azm\'any P. s\'et\'any 1/A H-1117 Budapest, Hungary \\
        E-mail: \email{pittler@bodri.elte.hu}}

\author{Laszlo Ujfalusi \\
        Department of Theoretical Physics, Budapest University of Technology
        and Economics \\ H-1521, Budapest, Hungary}

\author{Imre Varga \\
        Department of Theoretical Physics, Budapest University of Technology
        and Economics \\ H-1521, Budapest, Hungary}

\abstract{Recently we found an Anderson-type localization-delocalization
  transition in the QCD Dirac spectrum at high temperature. Using spectral
  statistics we obtained a critical exponent compatible with that of the
  corresponding Anderson model. Here we study the spatial structure of the
  eigenmodes both in the localized and the transition region. Based on
  previous studies in the Anderson model, at the critical point, the
  eigenmodes are expected to have a scale invariant multifractal structure. We
  verify the scale invariance of Dirac eigenmodes at the critical point.}

\FullConference{The 32nd International Symposium on Lattice Field Theory,\\
		23-28 June, 2014\\
		Columbia University New York, NY}

\begin{document}

\section{Introduction}

\begin{figure}[bc!]
\begin{center}
\parbox{0.24\textwidth}{\mygraph{0.24\textwidth}{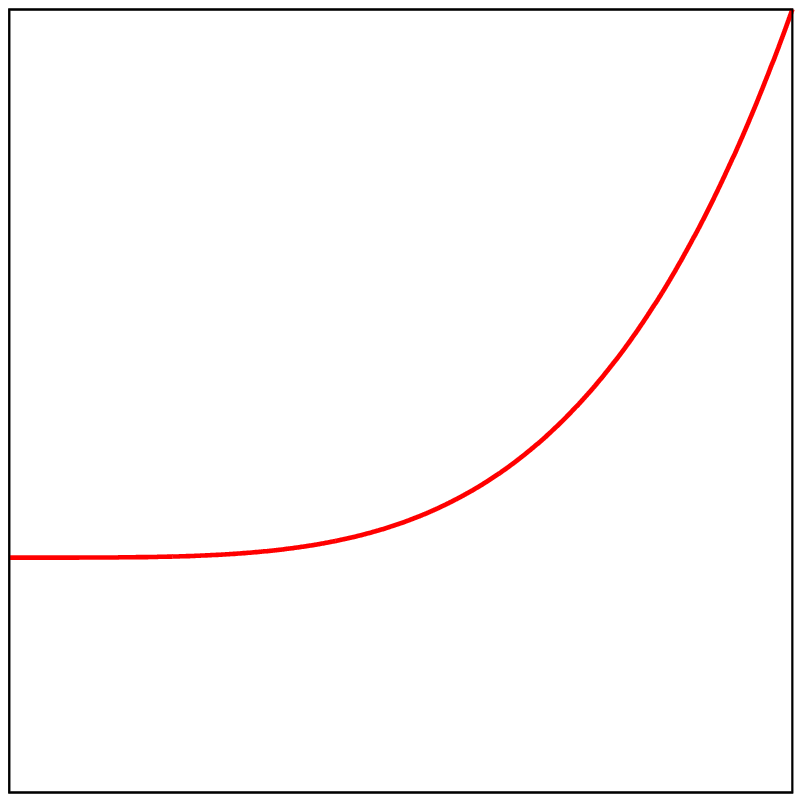} \\ 
                                                $\mbox{\hspace{3ex}} T<T_c$}
\parbox{0.24\textwidth}{\mygraph{0.24\textwidth}{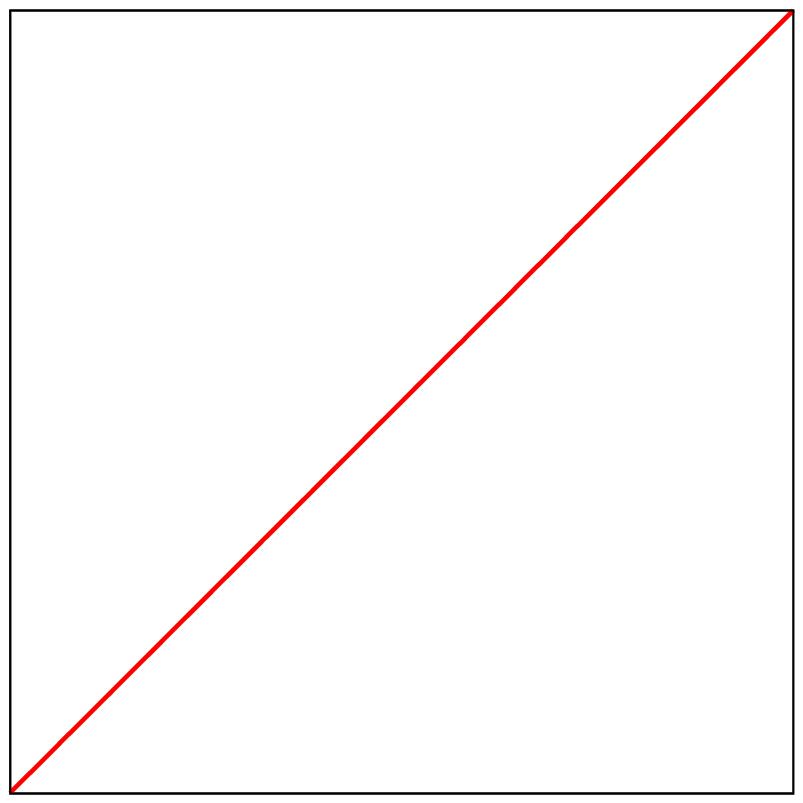} \\ 
                                          $\mbox{\hspace{3ex}}T\approx T_c$} 
\parbox{0.24\textwidth}{\mygraph{0.24\textwidth}{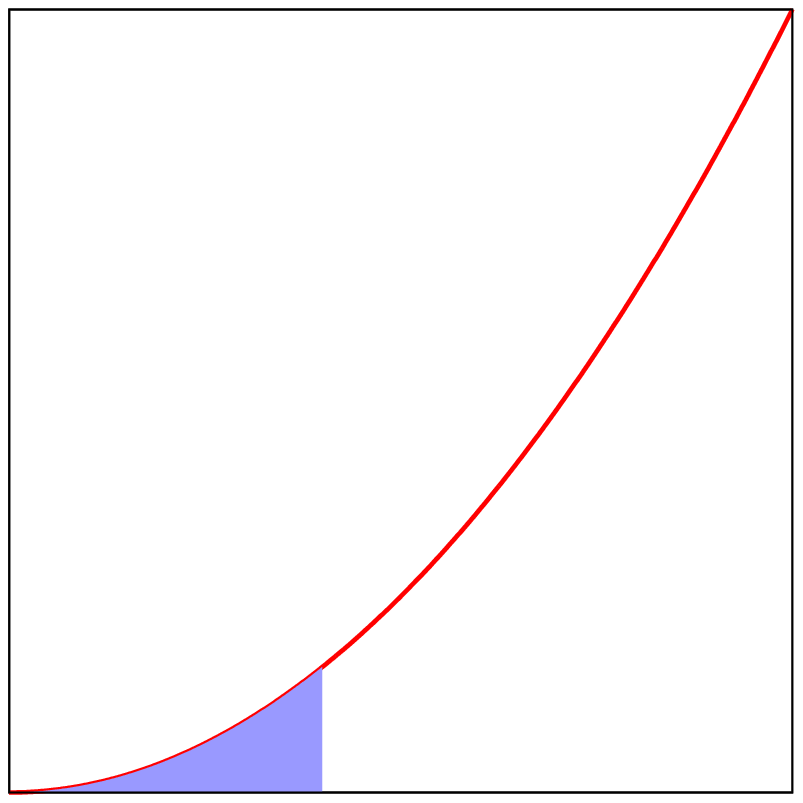} \\ 
                                                 $\mbox{\hspace{3ex}}T>T_c$} 
\parbox{0.24\textwidth}{\mygraph{0.24\textwidth}{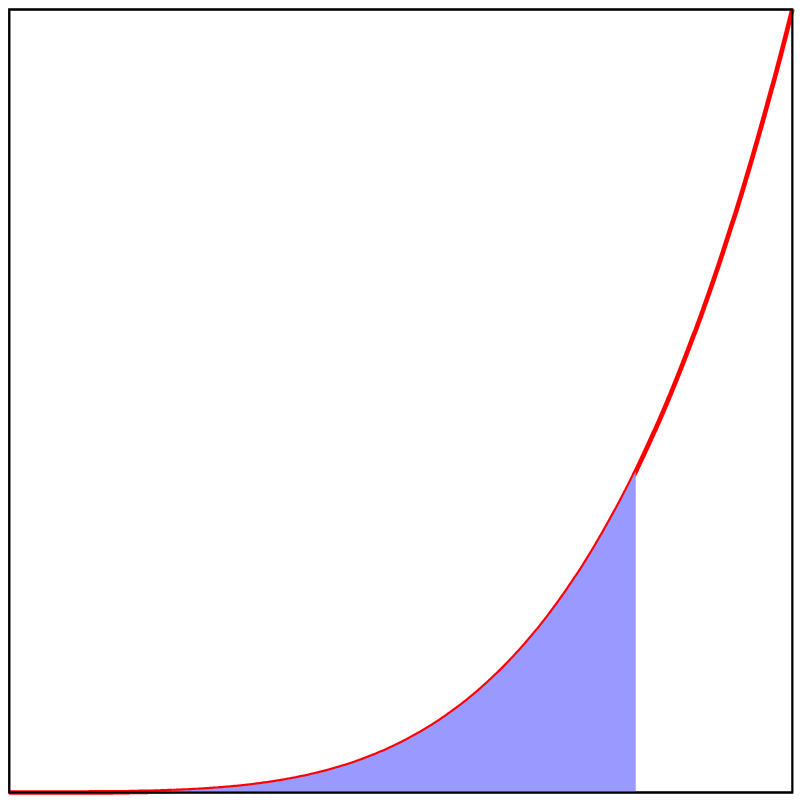} \\
  \parbox{18ex}{{\scriptsize \Blue{\hspace{9.5ex} localized}}
                 \Red{$\uparrow$} \\  
         \scriptsize \Red{\hspace{9.5ex}mobility edge}} }
\caption{\label{fig:Dspec_schem} Schematic picture of the spectral density of
  the Dirac operator around zero as the system crosses the critical
  temperature from below. The temperature increases from left to right and the
  shaded areas indicate the localized modes at the low-end of the spectrum.}
\end{center}
\end{figure}

It is well-known that the chiral cross-over of QCD is accompanied by a drastic
change in the low-end of the spectrum of the Dirac operator. In a theory with
massless quarks the spectral density at zero is proportional to the order
parameter of the phase transition from the chirally broken to the symmetric
phase \cite{Banks:1979yr}. Although in the real world even the lightest quarks
are massive, most of the chiral symmetry breaking in QCD at low temperature
still comes from spontaneous breaking. As a result, the qualitative picture of
the Dirac spectrum across the transition is similar to that in the massless
(chiral) limit. In Fig.~\ref{fig:Dspec_schem} we show a cartoon of how the
spectral density changes in QCD across the transition. At low temperature the
spectral density at zero is finite, signaling the spontaneous breaking of
chiral symmetry. At a point around the critical temperature of the cross-over,
the spectral density at zero vanishes and the low-end of the spectrum becomes
more and more sparse as the temperature increases further.

These features of the Dirac spectrum are rather well-known. What is not so
widely known, however, is that across the transition not only the spectral
density but also the physical nature of the low eigenmodes changes
considerably. Below the critical temperature $T_c$ the lowest modes are
delocalized in the whole volume, no matter how large that is.  In contrast,
above $T_c$ the lowest part of the Dirac spectrum consists of modes localized
on the scale of the inverse temperature \cite{Kovacs:2012zq}. However, even at
high temperatures, the modes farther away from the edge, in the bulk of the
spectrum, are still delocalized. In Fig.~\ref{fig:Dspec_schem} the shaded
regions indicate the localized modes, that are separated from the delocalized
ones by the so called {\it mobility edge}. The localization properties of the
eigenmodes is also reflected in the spectral statistics: eigenvalues
corresponding to localized modes obey Poisson statistics, those corresponding
to delocalized modes are described by Wigner-Dyson statistics, known from
random matrix theory, and widely used in the study of Dirac spectra below
$T_c$.

The transition in the spectrum from localized to delocalized modes is
analogous to Anderson transitions, first proposed to take place in conductors,
in the presence of disorder in the crystal lattice \cite{Anderson58}. In fact,
already ten years ago it was suggested that the QCD chiral transition might be
understood as an Anderson transition \cite{GarciaGarcia:2004hi}. Later, using
calculations in the instanton liquid model \cite{GarciaGarcia:2005vj} and
lattice QCD \cite{GarciaGarcia:2006gr}, qualitative support was obtained for
this picture. Subsequently some of us studied Dirac spectra well above
$T_c$. We found that the lowest part of the spectrum is always localized and
that the position of the mobility edge, separating localized and delocalized
modes, is controlled by the temperature. Most importantly, the temperature
where the mobility edge goes to zero and localized modes disappear, coincides
with the chiral and deconfining cross-over temperature
\cite{Kovacs:2012zq}. This lends further support to the conjecture that the
Anderson-type transition in the spectrum and the chiral transition at $T_c$
are strongly related. In the present work, to elucidate this connection
further, we propose to study the spatial structure of the Dirac eigenmodes
across the transition. In particular, we present preliminary results
supporting the expectation that at the transition the Dirac eigenmodes show
signs of critical behavior similar to those found in Anderson transitions.

\section{Anderson transition in the Dirac spectrum above $T_c$}

The connection between Anderson transitions and the transition in the Dirac
spectrum above $T_c$ is much more than a loose analogy. In previous work,
using finite size scaling of the unfolded level spacing distribution, we
showed that in the thermodynamic limit at the critical point $\lambda_c$ (the
mobility edge) in the spectrum, spectral properties change in a non-analytic
way. This also implies a diverging correlation length in the
eigenmodes. Moreover, we found that the critical exponent $\nu$ characterizing
this singularity is compatible with that of the three-dimensional unitary
Anderson model \cite{Giordano:2013taa}. This strongly suggests that above
$T_c$ there is indeed a genuine Anderson transition from localized to
delocalized modes in the Dirac spectrum and this transition is in the same
universality class as that of the Anderson model.

However, we immediately have to point out that this ``phase transition'' in
the spectrum does not imply a physical phase transition in QCD. This is
because there is no physical control parameter that could tune the system to
the critical point $\lambda_c$. The mobility edge $\lambda_c$ is just a point
in the spectrum of the Dirac operator and thermodynamic quantities are
averages over the whole Dirac spectrum. This has to be contrasted with the
situation in Anderson transitions occurring in condensed matter systems. In
that case the mobility edge is a genuine physical energy, a point in the
spectrum of the one-electron Hamilton operator. By changing some physical
control parameter (electron density, external field) the Fermi energy can be
driven through the mobility edge. As the Fermi energy passes from the
delocalized (conducting) states to the localized (non-conducting) states, the
zero temperature conductivity changes non-analytically and the Anderson
transition implies a genuine physical phase transition.

\begin{figure}
\begin{center}
\includegraphics[width=0.8\columnwidth,keepaspectratio]{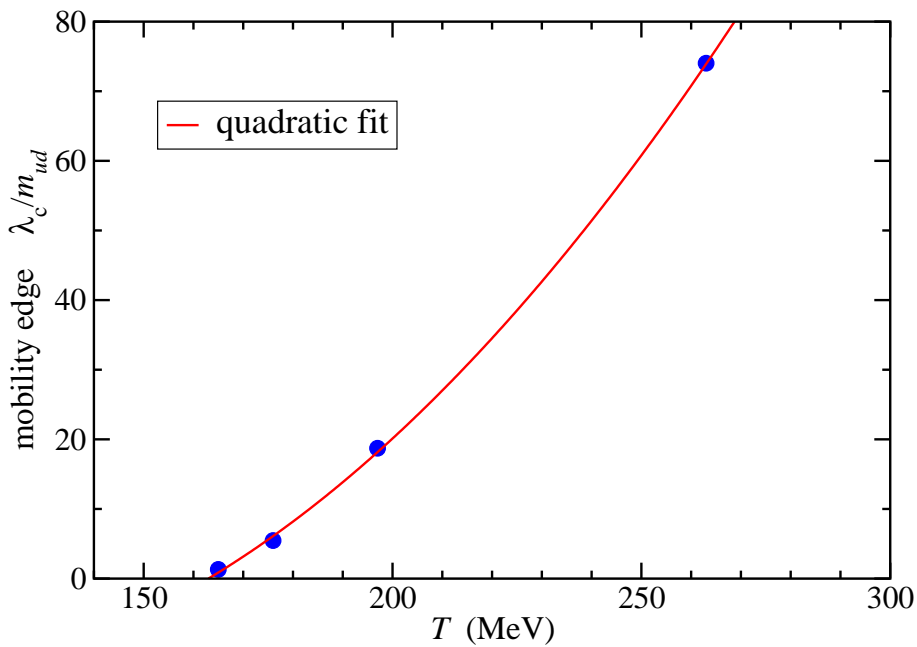}
\caption{\label{fig:lc_vs_T} The temperature dependence of the mobility edge
  $\lambda_c$ obtained from lattice QCD simulations with 2+1 flavors of
  staggered quarks at the physical point. }
\end{center}
\end{figure}

This is an important difference between the Anderson transitions in condensed
matter systems and QCD. However, even in QCD there is a possibility for the
Anderson transition to be accompanied by a real phase transition. This is at
the temperature where the mobility edge reaches zero when the temperature is
lowered from above $T_c$. In the thermodynamic limit, this is expected to
happen at a well-defined temperature that we call $T\msub{cAnd}$. Note that
$T\msub{cAnd}$ is well-defined even if the chiral transition is only a
cross-over, like in QCD. To demonstrate how the mobility edge goes to zero, in
Fig.~\ref{fig:lc_vs_T} we plot the temperature dependence of the mobility edge
in QCD with $N_f=2+1$ light staggered quark flavors with physical masses. We
normalized the mobility edge by dividing it by the bare light quark mass to
obtain a quantity that has a well-defined continuum limit. This also made it
possible to plot data from simulations with different lattice spacings in the
same figure. The line is a quadratic fit to the data. An extrapolation yields
$T\msub{cAnd}=163(2)$MeV, which is within the range of the chiral and
deconfining cross-over \cite{Borsanyi:2010bp}.

We have seen that the finite temperature cross-over and the (dis)appearance of
localized modes happen around the same temperature. In some QCD-like models
with a genuine chiral phase transition not only $T\msub{cAnd}$ but $T_c$ is
also well-defined and one can make a stronger statement. We found that in that
case the two critical temperatures coincide (see Pittler's contribution at
this conference \cite{Pittler_lat14}). It would be interesting to make
connections between the physical quantities characterizing the critical
behavior at the two transitions. To this end, we would need a detailed study
around the point in the ``phase diagram'' where the $\lambda_c(T)$ critical
line reaches the horizontal axis. The crucial question is whether the vanishing
of the mobility edge is accompanied by an abrupt change in the spectral
density in a finite interval around zero virtuality. If this happens and the
change in the spectral density is non-analytic in the temperature, the theory
has a phase transition.

\section{Dirac eigenmodes at the QCD Anderson transition}

Previously, most of our quantitative results about the QCD Anderson transition
were based on statistical properties of the Dirac spectrum. However, it is
well known in the case of Anderson transitions that the spatial structure of
eigenmodes also encodes useful information about the transition. In the
Anderson model, exactly at the critical point, the eigenmodes develop a
peculiar multifractal structure \cite{Evers-Mirlin-RMP}. Recently this has
been exploited for a high precision determination of the critical exponent
using finite size scaling of wave function observables \cite{Multifrac-FSS}.

Why is the structure of Dirac eigenmodes important in connection with the
chiral transition? The reason is that any fermionic observable can be
decomposed in terms of eigenmodes. In particular, the disconnected chiral
susceptibility can be written in terms of eigenmode correlators of the form
\be
 G(x) =  \langle |\psi(0)|^2 \, |\psi(x)|^2 \rangle.
\ee
If the eigenmodes have a (multi)fractal structure then the behavior of the
correlator $G(x)$ is closely related to the fractal dimension(s)
characterizing the eigenmodes \cite{Nakayam-Yakubo}. This can provide a link
between fractal properties of the critical eigenmodes and (pseudo)-critical
properties of the chiral transition.

Before outlining our proposal for quantities to look at in connection with the
eigenmodes, we have to explain some basic concepts concerning
multifractals. Let us assume that $f$ is a function $f: \mathbb{R}^d
\rightarrow \mathbb{R}$. Let us define the level sets of the function as 
\be
  S(a)=\{x: a<f(x)<a+\Delta a \}.
\ee
If $f$ is a smooth function then the level sets either have dimension $d$ or
are empty, depending on whether the given neighborhood of $a$ has an
intersection with the range of $f$ or not. In contrast, if the dimension of
the level sets $d(a)$ depends non-trivially on $a$, we call $f$ a multifractal
weight function. We would like to study the average properties of a collection
of such functions $f(x)=|\psi(x)|^2$ that are given only at the lattice sites
and normalized to unity. We subdivide the lattice of linear size $L$ into
smaller boxes of linear size $l\ll L$ and compute the coarse grained box weights
\be
    \mu_k(l)=\sum\msub{x $\in$ box k} |\psi(x)|^2,
     \label{eq:mu}
\ee
where the sum runs over those lattice sites that are contained in box $k$. If
the eigenmode has a multifractal structure, the system is scale invariant and
consequently the distribution of box weights is expected to depend
only on the ratio $l/L$ but not on $l$ and $L$ separately. If this is the case
then the $l/L$ dependence of the distribution can be used to obtain the
multifractal dimensions characterizing the eigenmodes \cite{Multifrac-FSS}.

\begin{figure}
\begin{center}
\psfrag{lambdaa}[0.0]{{\hspace{7ex} \Large $\lambda a$}}
\psfrag{logmuavg}[270.0]{{\Large \hspace{7ex} $\alpha_1$}}
\psfrag{leq3fm}[0.0]{{\hspace{7ex} \large $L=3$fm}}
\psfrag{leq4fm}[0.0]{{\hspace{7ex} \large $L=4$fm}}
\psfrag{leq5fm}[0.0]{{\hspace{7ex} \large $L=5$fm}}
\psfrag{leq6fm}[0.0]{{\hspace{7ex} \large $L=6$fm}}
\psfrag{leq7fm}[0.0]{{\hspace{7ex} \large $L=7$fm}}
\includegraphics[width=0.8\columnwidth,keepaspectratio]{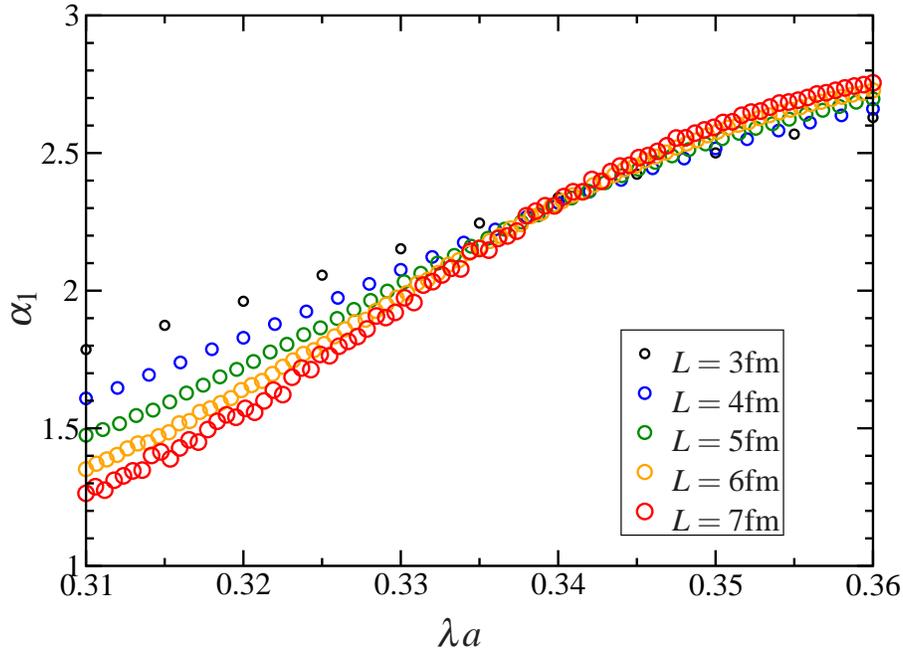}
\caption{\label{fig:alpha} The information dimension $\alpha_1$ of eigenmodes
  as a function of the location in the spectrum. The average was computed
  separately for eigenmodes located in different narrow windows in the
  spectrum. The various symbols represent data taken in systems of different
  spatial sizes while keeping the ratio of coarse graining box size to the
  system size fixed at $l/L=0.125$.}
\end{center}
\end{figure}

As a first check, we would like to verify scale invariance at the mobility
edge using a quantity derived from the box weights of eq.~(\ref{eq:mu}).  For
illustration, here we look at a quantity that has a transparent physical
interpretation, called the information dimension. It is defined as
\be
   \alpha_1 = \frac{1}{\log(l/L)} \langle \sum_k \mu_k \log(\mu_k) \rangle, 
\ee
where the sum runs over all the boxes of size $l$ and the averaging is
performed for eigenmodes in a narrow spectral window on several gauge
configurations. Since the eigenmodes are normalized, the quantity that is
averaged is the information entropy of a given eigenmode corresponding to the
given coarse graining box size $l$. It can be easily verified that for
eigenmodes uniformly spread in the whole system volume, $\alpha_1$ is equal to
the dimension of the system. On the other hand, for localized modes, the
information entropy goes to zero if $L\rightarrow \infty$ and $l/L$ is kept
fixed. This is because as the coarse graining box size becomes larger,
eventually the whole eigenmode will be typically contained in one box.

The eigenmodes for the computation of $\alpha_1$ were obtained from lattice
QCD simulations performed with the action of Ref.~\cite{Aoki:2005vt} at a fixed
temperature well above $T_c$.  We used spatial system sizes in the range
$L=3-7$~fm. In Fig.~\ref{fig:alpha} we plot $\alpha_1$ as a function of the
location of the eigenmodes in the spectrum. The different symbols correspond
to different spatial system sizes and the coarse graining box size $l$ was
always chosen such that the ratio $l/L$ was kept fixed at $l/L=0.125$. In the
figure we can see that around $\lambda a=0.34$ the information dimension
becomes independent of the system size, indicating that this is the critical
point (mobility edge) in the spectrum where eigenmodes are scale invariant. It
is reassuring that the critical point obtained here from eigenmode properties
agrees with the one that we computed previously from spectral statistics
\cite{Giordano:2013taa}.

Using quantities like the above described information dimension, the finite
size scaling analysis of Ref.~\cite{Giordano:2013taa} can be repeated. This is
work in progress but our preliminary results already indicate that both the
critical point and the critical exponent we obtain are consistent with the
previously determined values that were based on spectral statistics. However,
it is interesting to note that the uncertainty of the results based on
eigenmode observables is larger than those obtained from spectral statistics
using the same ensemble of lattice configurations. This is because scaling
violations, due to operators irrelevant in the RG sense, appear to be more
sizeable for eigenmode observables than for observables derived from spectral
statistics. For better precision one might need to consider systems of larger
spatial sizes.

\section{Conclusions}

Previously we found an Anderson-type localization-delocalization transition in
the spectrum of the QCD Dirac operator at high temperature. Based on a finite
size scaling analysis of spectral observables we obtained strong evidence that
the transition is in the same universality class as that of the corresponding
(3d unitary) Anderson model. In the present paper we proposed a quantitative
study of the spatial structure of the corresponding eigenmodes. This could
provide further information concerning the nature of the transition. As a
first consistency check we verified that at the critical point in the spectrum
the coarse grained eigenmode box probabilities become scale invariant, as
expected in Anderson transitions. We found that the critical point obtained
from the scale invariance of the eigenmodes coincides with the one previously
determined from spectral observables. Preliminary results of a finite size
scaling analysis of eigenmode data indicate that the resulting critical
exponent is also consistent with the one obtained from spectral data. This
shows that in QCD, spectral and eigenmode properties signal the Anderson
transition in a consistent way, exactly like in the Anderson model. It would
be also interesting to determine the dimensions characterizing the
multifractal eigenmodes at the critical point and compare those to results
from the Anderson model. To this end we intend to study the dependence of the
box probabilities on the ratio of the box size and the system size ($l/L$). In
the present study this quantity was kept fixed.


\begin{thebibliography}{99}

\bibitem{Banks:1979yr} 
  T.~Banks and A.~Casher,
  Nucl.\ Phys.\ B {\bf 169}, 103 (1980).

\bibitem{Kovacs:2012zq} 
  T.~G.~Kovacs and F.~Pittler,
  Phys.\ Rev.\ D {\bf 86}, 114515 (2012)
  [arXiv:1208.3475 [hep-lat]].

\bibitem{Anderson58}
P.~W.~Anderson, Phys.\ Rev.\ {\bf 109}, 1492 (1958).

\bibitem{GarciaGarcia:2004hi} 
  A.~M.~Garcia-Garcia and K.~Takahashi,
  Nucl.\ Phys.\ B {\bf 700}, 361 (2004)
  [cond-mat/0403557].

\bibitem{GarciaGarcia:2005vj} 
  A.~M.~Garcia-Garcia and J.~C.~Osborn,
  Nucl.\ Phys.\ A {\bf 770}, 141 (2006)
  [hep-lat/0512025].

\bibitem{GarciaGarcia:2006gr} 
  A.~M.~Garcia-Garcia and J.~C.~Osborn,
  Phys.\ Rev.\ D {\bf 75}, 034503 (2007)
  [hep-lat/0611019].

\bibitem{Giordano:2013taa} 
  M.~Giordano, T.~G.~Kovacs and F.~Pittler,
  Phys.\ Rev.\ Lett.\  {\bf 112}, no. 10, 102002 (2014)
  [arXiv:1312.1179 [hep-lat]].

\bibitem{Borsanyi:2010bp} 
  S.~Borsanyi {\it et al.}  [Wuppertal-Budapest Collaboration],
  JHEP {\bf 1009}, 073 (2010)
  [arXiv:1005.3508 [hep-lat]].

\bibitem{Pittler_lat14} 
  M.~Giordano, S.~D.~Katz, T.~G.~Kovacs and F.~Pittler,
   PoS LATTICE {\bf 2014}, 214 (2014).

\bibitem{Evers-Mirlin-RMP}
  F.~Evers and A.~D.~Mirlin, 
  Rev.\ Mod.\ Phys.\ {\bf 80}, 1355 (2008). 

\bibitem{Multifrac-FSS}
  A.~Rodriguez, L.~J.~Vasquez, K.~Slevin and R.~R\"omer,
  Phys.\ Rev.\ Lett.\ {\bf 105}, 046403 (2010); Phys.\ Rev.\ B {\bf 84},
  134209 (2011).

\bibitem{Aoki:2005vt} 
  Y.~Aoki, Z.~Fodor, S.~D.~Katz and K.~K.~Szabo,
  JHEP {\bf 0601}, 089 (2006)
  [hep-lat/0510084].

\bibitem{Nakayam-Yakubo}
  T.~Nakayama and K.~Yakubo, ``Fractal concepts in condensed matter physics'',
  1st ed.\ Springer Verlag (2003), p.\ 12, 48-49.



\end{thebibliography}
\end{document}